\documentclass{article}

\usepackage{arxiv}

\usepackage[utf8]{inputenc} 
\usepackage[T1]{fontenc}    
\usepackage{hyperref}       
\usepackage{url}            
\usepackage{booktabs}       
\usepackage{amsfonts}       
\usepackage{nicefrac}       
\usepackage{microtype}      
\usepackage{graphicx}
\usepackage{doi}
\usepackage{setspace}
\usepackage{changes}
\usepackage{amsmath}
\usepackage{float}
\usepackage{xcolor}
\usepackage{bm}
\usepackage{color,soul}

\usepackage{lineno}
\title{\vspace{-1cm} Imaging concentration fields in microfluidic fuel cells as a mass transfer characterization platform}
\date{\textit{Revised on \today}}

\author{Marine Garcia \\
	Arts et Métiers Institute of Technology, CNRS, Université de Bordeaux, Bordeaux INP\\
	Institut de Mécanique et d'Ingénierie (I2M), Bâtiment A11,\\
	351 Cours de la Libération, 33405 Talence, France \\	
	 \AND
	 Alain Sommier \\
	 CNRS, Arts et Métiers Institute of Technology, Université de Bordeaux, Bordeaux INP \\	 
	Institut de Mécanique et d'Ingénierie (I2M), Bâtiment A11,\\
	351 Cours de la Libération, 33405 Talence, France \\	 
	\AND
	 Dominique Micheau \\
     Université de Bordeaux, CNRS, Bordeaux INP,\\
	 Institut de Chimie de la Matière Condensée de Bordeaux (ICMCB)\\	
	 F-33600 Pessac, France \\
	 \AND
	 Gérald Clisson \\
     CNRS, Solvay, LOF, UMR 5258\\ 
	178, avenue du Docteur Schweitzer\\
	33608 Pessac, France\\	
	 \And
	 Jean-Christophe Batsale \\
	 Arts et Métiers Institute of Technology, CNRS, Université de Bordeaux, Bordeaux INP, \\
	Institut de Mécanique et d'Ingénierie (I2M), Bâtiment A11,\\
	351 Cours de la Libération, 33405 Talence, France \\	
	 \And
	 \href{https://orcid.org/0000-0002-4614-1867}{\includegraphics[scale=0.06]{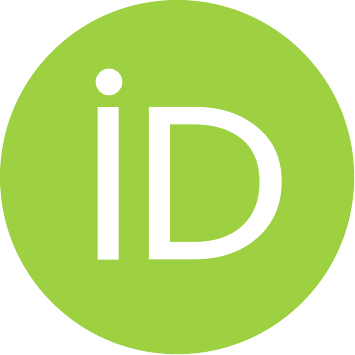}\hspace{1mm}Stéphane Chevalier}\thanks{Corresponding Author: Prof. Stéphane Chevalier, \texttt{stephane.chevalier@u-bordeaux.fr}} \\
	Arts et Métiers Institute of Technology, CNRS, Université de Bordeaux, Bordeaux INP\\
	Institut de Mécanique et d'Ingénierie (I2M), Bâtiment A11,\\
	351 Cours de la Libération, 33405 Talence \\
}

\renewcommand{\shorttitle}{\textit{Submitted to Electrochimica Acta.}}

\hypersetup{
pdftitle={Concentration field imaging in microfluidic fuel cells based on operando visible spectroscopy},
pdfsubject={orignal research paper},
pdfauthor={S. Chevalier},
pdfkeywords={Microfluidic channel, Electrochemical reaction, Mass diffusion ,  Current density distribution , Analytical model , Fuel cells},
}
\begin{document}
\maketitle
\vspace{2cm}
\begin{abstract}
Microfluidic fuel cells (MFCs) are microfluidic electrochemical conversion devices that are used to power small pieces of electrical equipment. Their performance relies on the improvement of the mass transfer of the reactants at the electrode interface. {In this work, a MFC is developed to implement a novel imaging technique that allows the measurement of reactant concentration fields, featuring formic acid as the fuel and potassium permanganate as the oxidant. The concentration fields were imaged } based on transmitted visible spectroscopy, which links the light intensity passing through the MFC to its local reactant concentration. {An analytical model was developed to estimate the mass diffusivity and kinetic reaction rate coefficient. For the first time, mass transport and transfer coefficient were simultaneously measured during operation. These parameters estimated using the proposed technique can be implemented in a numerical model to predict the MFC performance and concentration distribution. This work paves the way toward advanced imaging tools for operando mass transfer characterizations in microfluidics and Tafel kinetic characterization in many electrochemical devices.}

\end{abstract}

\vspace{2cm}

\clearpage

%
%
%
%

\section{Introduction}
\medskip
Microfluidic fuel cells (MFCs) are {microscale systems} used to convert the chemical energy contained in fuels directly into electricity \cite{Kjeang2009,Lee2013}, making these devices promising energy sources. Such devices are composed of a microfluidic channel that ensures rather good control of the hydrodynamic conditions. {In the channel, two electrodes are embedded to enable a oxidation reaction at the anode and a reduction reaction at the cathode}. Such an MFC can be used both {as fuel cell or electrolyzers}, which makes this technology a promising candidate for energy conversion and storage. A wide variety of MFCs are present in the literature, and more details about them can be found in the following comprehensive reviews \cite{Ibrahim2022,Zhou2021,Wang2021}. The present study focuses on a coflow membraneless MFC \cite{Choban2004} using formic acid (HCOOH) and potassium permanganate (KMnO$_4$), as this system is relatively robust, is compatible with classical soft photolithography micro fabrication techniques, uses nonhazardous chemicals and is easy to operate \cite{Lopez-Montesinos2011,Salloum2008}.\\

Three main phenomena govern  MFC performance. First, the mass transfer is based on the diffusion, advection and reaction of the chemical species in the microchannel. The second phenomenon is linked to charge transfer in the electrolyte and the electrodes, several authors have reported a thorough description of such \cite{Bazant2004,Chevalier2021}. The last phenomenon is the entropy generated during the energy conversion, which is transformed into heat and impacts the rate of mass diffusion and the electrochemical kinetics. Thus, optimal MFC performance relies on accurate control and knowledge of these phenomena {which have the potential to be characterize through operando contactless imaging technique.} \\

Over the last few decades, a large number of MFC numerical models were developed to predict their performance \cite{Al-Fetlawi2010,Bazylak2005,Shah2010,Zhi2021,Esan2020}. Among them, Gervais \& Jensen \cite{Gervais2006} describe several mass transport and electrochemical reaction modelling methods. These models rely on the knowledge of the mass diffusivity and kinetic reaction rate coefficients. Multiphysical modelling is also often used, such as in the work of Wang et al. \cite{Wang2018}. Such models need many important parameters, which are difficult to find in the literature or to measure ex situ. {In the literature, numerical studies are mainly compared to experimental results using the polarization curves. The comparison can be improved by studying the mass transfer that occurs in an operating cell. Using imaging techniques, in-situ characterization of the mass transport enables the derivation of the mass diffusivity coefficient and the kinetic reaction rate, allowing the development of more accurate models.} \\

Several studies in the literature have shown great interest in characterizing mass transport using imaging methods \cite{Lee2015,Sun2007,Jindal2017,Liu2019}. For example, Sun et al. \cite{Sun2007} used an optical microscope to study permanganate diffusion into formic acid. Their work was mainly qualitative to illustrate the hydrodynamic flow in their MFC. Lu et al. \cite{Lu2018} used optical and fluorescence imaging techniques to validate their MFC model and to measure the concentration distribution profile at one position in the channel. However, imaging the concentration field in the microchannel has yet to be been reported. More advanced imaging techniques based on infrared techniques \cite{Chan2012,Ryu2017,Chevalier2021b,Perro2016} or X-ray \cite{Watanabe2020} can also be used to characterize the MFC structures, concentration fields or two-phase flow distribution. Among them, visible or ultraviolet (UV) spectroscopy are particularly efficient for measuring the chemical concentration of compounds in aqueous solutions, as light in these wavelengths are not strongly absorbed by water \cite{Yue2013}. However, although many imaging techniques are used for concentration measurements, thorough operando characterization of the mass transfer and the simultaneous acquisition of the electrochemical performance has yet to be implemented. Such a study would require designing a MFC, a specifically tailores imaging setup and a potentiostat to control the electrochemical conditions of the MFC. Although these challenges appear to be ambitious, the results that the operando images would yield are of prime interest for all MFC research. {In fact, }they would enable the measurement of the main parameters governing the cell performance such as the mass diffusivity, electrochemical kinetics and reactant concentration distribution.\\

The main goal of the present work is to report the use of imaging visible spectroscopy to measure the operando MFC mass transfer, specifically the mass diffusivity and kinetic reaction rate coefficient. Such a technique has already been used for various global concentration measurements in microfluidic reactors \cite{Rizkin2019}, but rarely to image a concentration field, despite it could be adapted for MFC transient concentration field characterization. {The physical properties and performance of a MFC can be estimated from a model of the hydrodynamic, mass transfer and electrochemical reaction.} In the first section, a description of the MFC design for visible spectroscopy, and the associated imaging setup is presented. This is followed by a description of the analytical model for concentration diffusion in the channel used to perform the parameter estimation. In the results section, the experimental polarization curve and concentration fields are presented and compared to the model. A thorough analysis of the technique is performed, and its limitations are discussed.

\section{Methods}

\subsection{Fabrication of MFC}
A chip featuring a T-shaped microfluidic channel was fabricated using standard photolithography. The microchannel height is 25 µm, width is 3 mm, and length is 15 mm. This specific aspect ratio was used to facilitate light transmission through the MFC and to reduce the transfers to 2D; see the model in the next section. A negative photoresist was spin coated on a silicon wafer, covered with a photomask and exposed to UV light. It was then submerged in a propylene glycol methyl ether acetate (PGMEA) solution for development. The obtained mold was placed in a Petri dish and coated with 5 mm of polydimethylsiloxane (PDMS). After being cured, the PDMS was peeled off the mould and hole-punched to create two inlets and one outlet. \\

\begin{figure}[H]
\centering
\includegraphics[scale=.35]{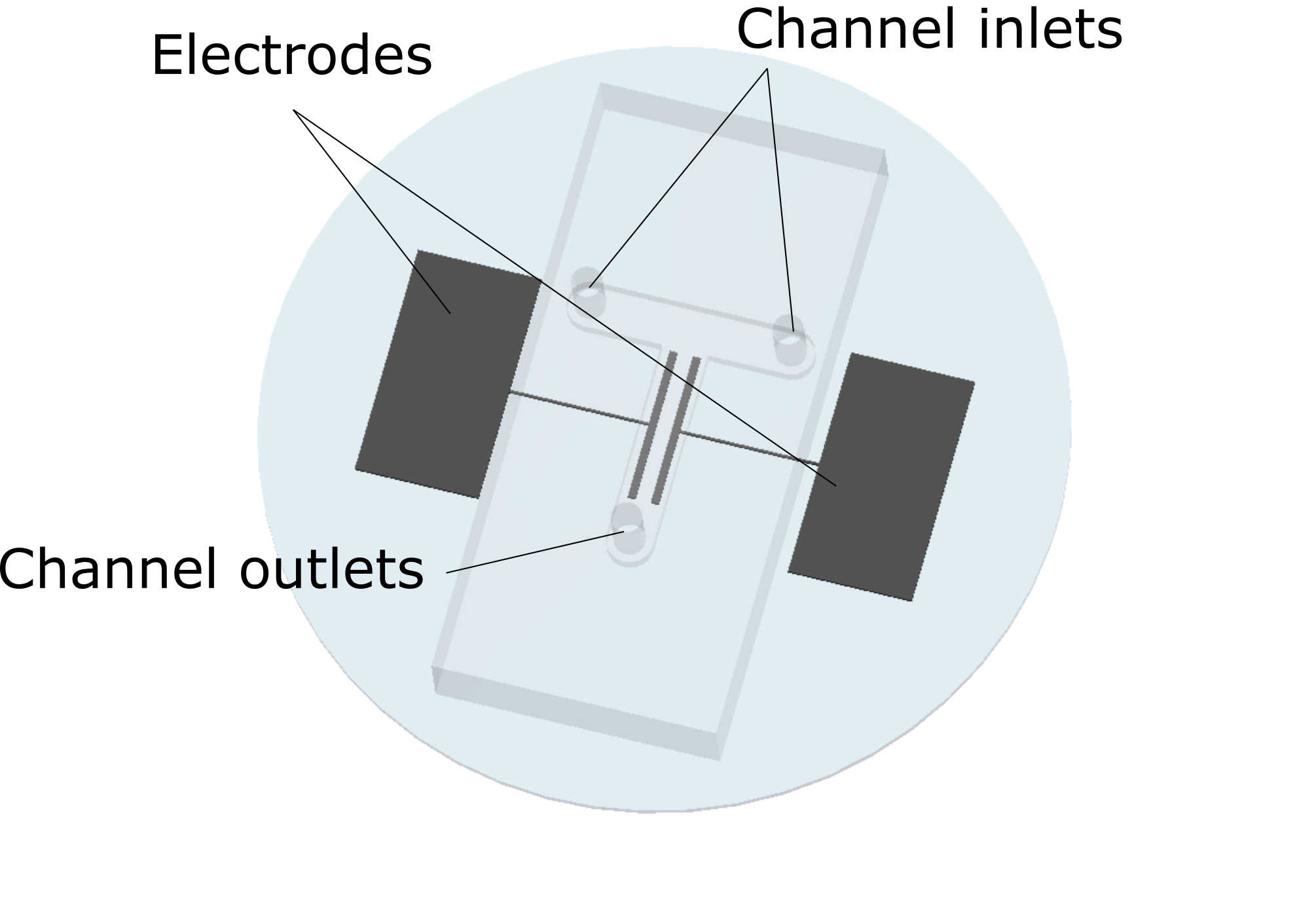}
\caption{Three-dimensional view of MFC.}
\label{f_MFC}
\end{figure}

For the electrodes, an inverse pattern was created on a glass wafer using the same photolithography process as for the PDMS stamp. The deposition was achieved by sputtering $\sim 60$ nm of titanium as an adhesion layer, and then subsequently $\sim 300$ nm of platinum for the catalyst material. The remaining photoresist was removed by submerging the wafer in a chemical etching solution {(MicropositT - MF-319)} to obtain only the platinum pattern directly on the glass substrate. The PDMS stamp was plasma activated and bound to the glass substrate, resulting in the complete MFC presented Figure 1. {The electrodes dimensions are 500} $\pm$ {3.5 µm} wide by 1  $\pm$ 0.0035 mm long, and are contained in the main channel where the reactants flow. More details and schematics of the microfabrication process can be found in the Supplementary Material.\\

\subsection{Experimental setup}
\begin{figure}[H]
\centering
\includegraphics[scale=.6]{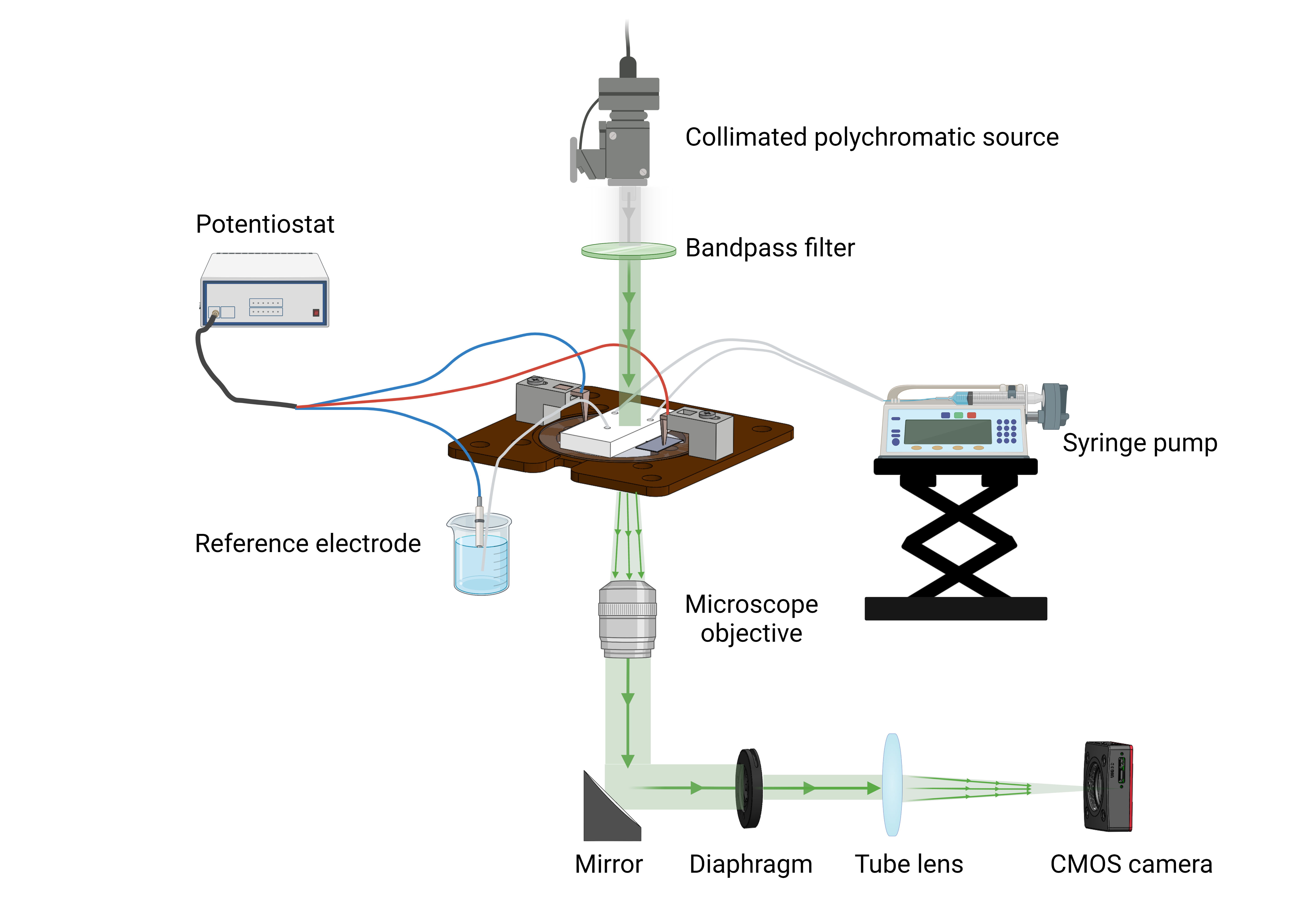}
\caption{Schematic of visible spectroscopic imaging setup used to measure in operando concentration.}
\label{f_setup}
\end{figure}

The concentration distribution and total current produced by the {MFC} were measured using the setup described in Figure \ref{f_setup}. It is made of a homemade inverse microscope. {The primary light source is a  white mounted LED (Thorlabs - MWW\textcolor{red}4) assembled with collimation adapter (Thorlabs -SM2F32-A) and placed 12 cm above the cell.} A narrow bandpass filter ($\lambda=540\pm 5 $ nm) is used to produce a monochromatic green light passing through the MFC. The light is finally collected through a microscope objective and a lens to produce an image  {with x1  magnification }on a CMOS camera (Zelux 1.6 MP Colour CMOS Camera).Only the green channel of the camera was used in the image postprocessing. The resulting spacial resolution is 3.45 µm/px leading to {observation field} of approximatively 5 by 4 mm.\\

The MFC is controlled using a potentiostat {(Biologic SP-300)} to measure the voltage and the current produced. {Electrical measurements are performed in a three electrodes configuration. An Ag/AgCl reference electrode is immersed in a beaker filled with 0.5M of sulfuric acid. The chemicals from the outlet of the chip are spilled in the beaker containing the reference electrode to ensure the electrical contact. In this configuration, anode and cathode potentials are measured simultaneously  in the same experiment allowing a full characterization.} The reactant flow rate is controlled using a syringe pump {(Cetoni Nemesys)} over a wide range from 0.5 to 100 µl/min.\\

The reactants (formic acid  and potassium permanganate) were chosen for good performance \cite{Kjeang2009}. In addition, permanganate potassium has the advantage of a clear absorption signature in the visible range {which allows the investigation of mass transport at the cathode. However, mass transport at the anode can not be studied since formic acid is transparent in the visible range.} At the anode, the formic acid oxidation is
\begin{equation}
HCOOH \longrightarrow CO_2+2H^++2e^-. 
\end{equation}
At the cathode, the  permanganate reduction is
\begin{equation}
MnO_4^-+8H^+ +5e^- \longrightarrow Mn^{2+}+4H_2O.
\label{e_manganese}
\end{equation}
In equation \ref{e_manganese}, when a current is produced, the permanganate ions (MnO$_4^-$) are transformed into manganese ions Mn$^{2+}$. Thus, the current applied through the MFC electrodes triggers a decrease in the permanganate concentration, which is measured by visible spectroscopy. {It is also assumed that Mn$^{2+}$ ions do not absorb light at the chosen wavelength, and that no CO$_2$ gas bubbles from the reaction appear during the experiment.}\\

The wavelength chosen in the imaging setup corresponds to the strongest light absorption of the MnO$_4^-$ ions, { whereas formic acid is completely transparent}. Thus, using the Beer-Lambert equation, one can link the variation in light intensity to the variation in permanganate concentration as
\begin{equation}
\Delta c_{exp} = -\kappa^{-1}\ln\left(\frac{I_0+\Delta I}{I_0}\right)\approx-\kappa^{-1}\left(\frac{\Delta I}{I_0}\right),
\label{e_BL}
\end{equation}
where $\kappa$ is the permanganate absorption coefficient (mM$^{-1}$), $I_0$ is the light intensity of the background and $\Delta I$ is the light intensity variation induced by the current production. The linearisation of the Beer-Lambert was used as the variation of light intensity is very small, i.e. less than 1.5\%. The permanganate absorption coefficient was measured to be $\kappa = 5,5\times 10^{-3}$ mM$^{-1}$ at 540 nm (see the calibration curve in Figure S3 in the Supplementary Material).

\subsection{Permanganate diffusion in the depletion zone}

\begin{figure}[H]
\centering
\includegraphics[scale=.7]{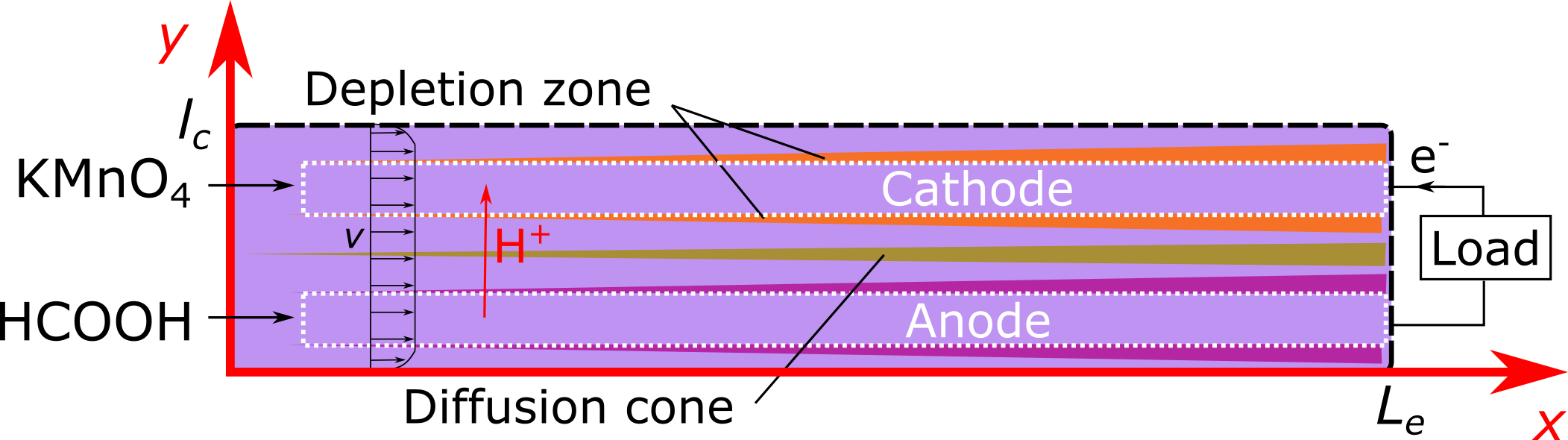}
\caption{2D schematic of channel with the electrodes placed at the bottom of the channel. The fluids flow from left to right. The main charge transport phenomena are indicated for sake of clarity, but are not modeled.}
\label{f_schema_3D}
\end{figure}

In our MFC geometry, the chemicals flow at a given velocity $v$, diffuse with a mass diffusivity $D$ and are consumed at the electrode interface at a rate given by the kinetic reaction rate coefficient $k_0$. The coefficient $k_0$ is linked to the current produced by the cell (see the Tafel law defined in appendix). The resulting concentration decreases at the electrode interface creating a depletion zone on each side of the electrodes where the concentration diffuses. Thus, the magnitude and the spatial distribution of this depletion zone enable the mass diffusivity $D$ and the kinetic reaction rate coefficient $k_0$ to be measure in the operating MFC.\\
Between the two reactants, a diffusion {zone} is also formed due to the interdiffusion between the HCOOH and the KMnO$_4$. If the velocity of the fluid is large enough, the diffusion {zone} should not interact with the depletion zone Figure \ref{f_schema_3D}.\\

Given the specific aspect ratio of the channel, i.e. height to width ratio larger than 100, all the mass transfer can be considered in 2D. {This assumption was checked numerically using COMSOL. The result presented in Section 3 off the supplementary material shows  an excellent agreement between the concentration profiles computed analytically and numerically is found, see Figure S5.} In addition, the velocity of the fluid can also be considered large enough to ensure a diffusion of the concentration using the semi-infinite assumption with no interaction with the diffusion {zone}. In this case, the diffusion of the permanganate in the depletion zone on each side of the electrodes can be analytically computed using a convolution product between the diffusive impulse response and the concentration at the electrode boundary, $y=e/2$, \cite{Crank1975} as
\begin{equation}
c(x,y) = \int_0^x c_e(x-x_0)\sqrt{\frac{\delta(y)}{\pi x_0^3}}\exp\left(-\frac{\delta(y)}{x_0}\right)dx_0, \forall y>e/2, \label{e_conv_prod}
\end{equation}
where $\delta=vy^2/(4D)$, $e$ is the electrode width (in the y-direction), and $c_e(x) = c(x,,y=e/2)$,  is the concentration at the channel/electrode interface. {The concentration at the interface} is mainly linked to the electrode dimension and the kinetic reaction rate constant, $k_0$. The calculation of this function is detailed in the appendix.\\
Equation \ref{e_conv_prod} is used to compute the concentration of the reactant in the depletion zone. The convolution product is computed using a numerical Laplace transform algorithm \cite{10.1145/361953.361969}. The analytical and relatively simple mathematical writing of this equation enables to use an inverse method to estimate $D$ and $k_0$.

\section{Results}
\subsection{Electrochemical performance of MFC}
\begin{figure}[H]
\centering
\includegraphics[scale=.7]{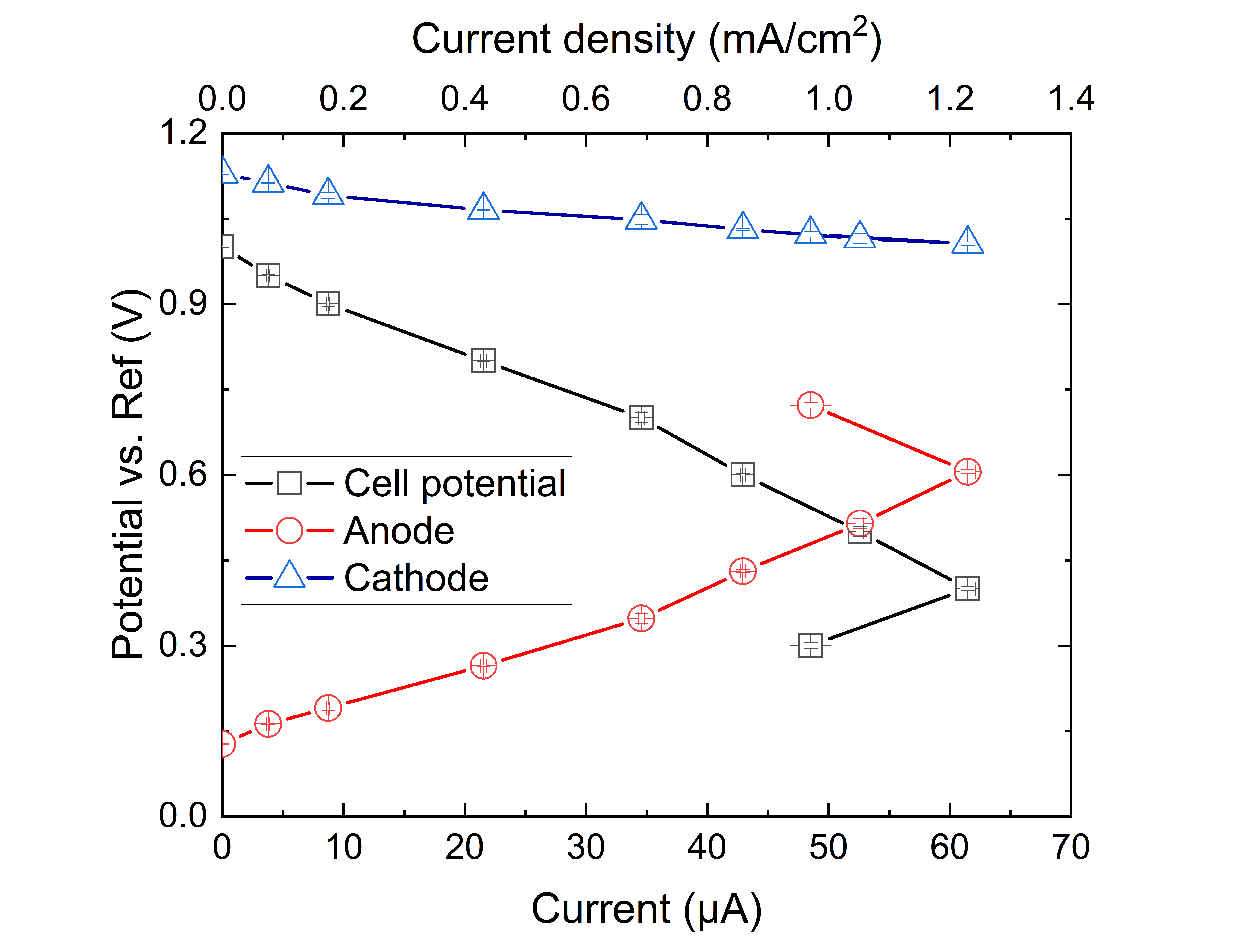}
\caption{MFC polarization curve. Each point is an average of the recorded data for one minute.}
\label{f_pol_curve}
\end{figure}

Before imaging the concentration field in the MFC, a polarization curve was measured. It was done using an aqueous solution with a {4M formic acid solution mixed with a 1M sulfuric acid solution in a ratio 1:1 at the anode}. At the cathode, an aqueous solution {containing 20mM of potassium permanganate mixed with 1M of sulfuric acid in a ratio 1:1 is used.} The flow rate was set to 5 µl/min for both inlets. The current and electrodes potentials were recorded for a range of cell potential between OCV and 0.2 V. Each cell potential was held for 5 min, and the current measured over the last min was average {in the reported} points. The electrode potential were measured against an Ag/AgCl reference electrode immersed in the MFC electrolyte.\\
The polarization curve obtained in Figure \ref{f_pol_curve} shows that the MFC underperformed compared to the literature \cite{Liu2019,Jayashree2005}. This is attributed to the catalyst used at the anode. For sake of simplifying in the fabrication process, the same catalyst was used for both the anode and the cathode, i.e. platinum. {However, using palladium as a catalyst could enhance the formic acid oxidation process as reported in the literature} \cite{Zhu2005,Antolini2009,Jiang2014}. The poor performance of the anode can be observed through the value of the anode potential, which increases drastically compared to the cathode potential. However, the present MFC was developed for imaging purposes and despite poor performance, the current density produced by the cell was enough to create a decreasing concentration gradient of the permanganate.

\subsection{Concentration field measurements}

\begin{figure}[H]
	\centering
	\includegraphics[scale=.5]{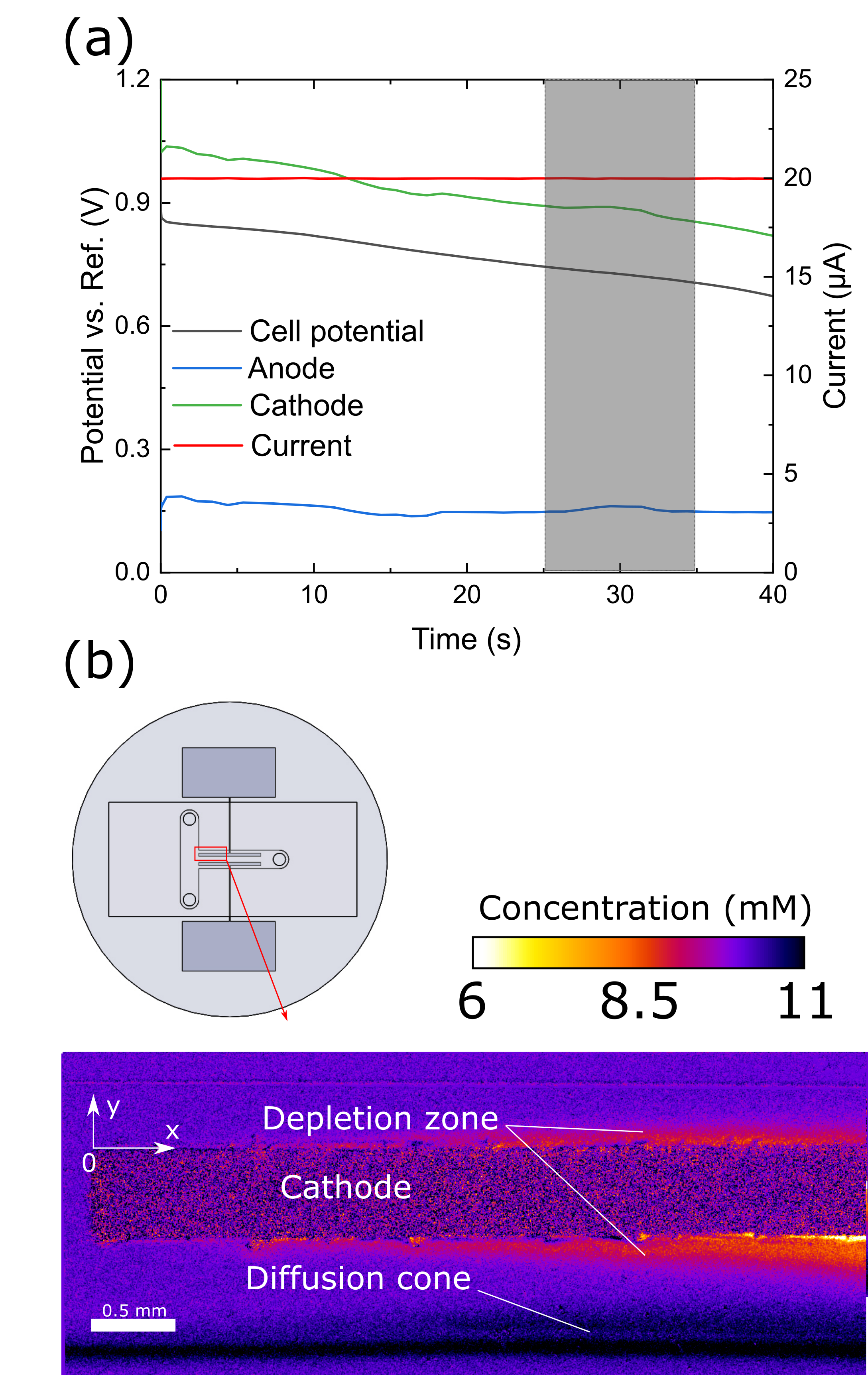}
	\caption{(a) MFC electrochemical parameter during the measurements of the concentration field at 20 µA. The grey rectangle indicates the time used to average the imaging data. (b) The steady state concentration field imaged at the inlet of the cathode (see the red rectangle in the MFC insert).}
	\label{f_fig5}
\end{figure}

The concentration field was imaged during a cell operation at {20 µA and 40 µA}. The images were acquired using the setup described in Figure \ref{f_setup}. The camera frame rate was set to 5 fps. The cell was imaged first at rest for a 15 s before generating a current to obtain the background image $I_0$. The anode and cathode flow rates were set to 1 µl/min, respectively. This flow rate ensures an average velocity of 0.42 mm/s and a residence time of 24 s which increase the {width} of the depletion zone on each side of the electrode. Consequently, the MFC can be considered as operating in steady state for any time longer than 24 s.\\
In Figure \ref{f_fig5}(a), the electrochemical performance of the cell during the imaging experiment at 20 µA is presented. The anode potential is quite steady, but a small decrease of the cathode potential is observed form 0.85 to 0.7 V. This behaviour is attributed to the creation of solid MnO$_2$ \cite{Wang2021,Salloum2008} which sediments on the electrode surface, lowering the cathode performance. Thus, the concentration field was imaged once the steady state is reached, i.e. after 25 s, for 10 seconds before an important solid MnO$_2$ layer covers the electrode. This period of time is indicated by the grey rectangle in Figure \ref{f_fig5}(a).\\

During this period of time, a change in light intensity of approximately 10-15 camera counts out of 900, i.e. less than 1.5\%, was detected on each side of the electrode. The signal-to-noise ratio (SNR) was estimated to be approximately {$\sim$3 }(noise is roughly 4/5 camera counts), which is very low. To reduce the signal to noise ratio, all images recorded during 10 s (50 images in total) were averaged and converted to concentration fields using equation \ref{e_BL}. Then, the absolute concentration field is deduced as $c_{exp}(x,y) = c_0-\Delta c_{exp}(x,y)$. This result is presented in Figure \ref{f_fig5}(b). {No signal can be recorded through the electrodes as they are fully opaque to visible light}. Figure \ref{f_fig5}(b), a concentration gradient appears and diffuses along the channel creating a depletion zone which is almost symmetrical on each side  the cathode. A slight change is visible at the bottom and can be explained by small flow instability in the MFC. The magnitude and the width of the depletion zone is more pronounced towards the end of the channel than at the inlet due to advection. Since the depletion zone are quite small, i.e. width of 300 µm at the maximum, a spatial resolution at the microscale was chosen (i.e. 3.45 µm/px, see section 2.2). { However, the field of view is limited, i.e. 5 mm in the channel direction, so only the first half of the channel is imaged(red rectangle in the Figure \ref{f_fig5}(b))}.\\

In the depletion zone, a maximum decrease of 2 mM is observed, which is a small variation in concentration. However, the results presented here demonstrated that it was possible to measure it using the rather simple setup proposed. From the data presented in Figure \ref{f_fig5}, a noise of $\pm 0.2$ mM is estimated. These results hig\textcolor{red}ight the sensitivity of the setup for detecting small concentration variations in MFCs.\\ 

\subsection{Estimation of $D$ and $k_0$}
\begin{figure}[H]
	\centering
	\includegraphics[scale=.5]{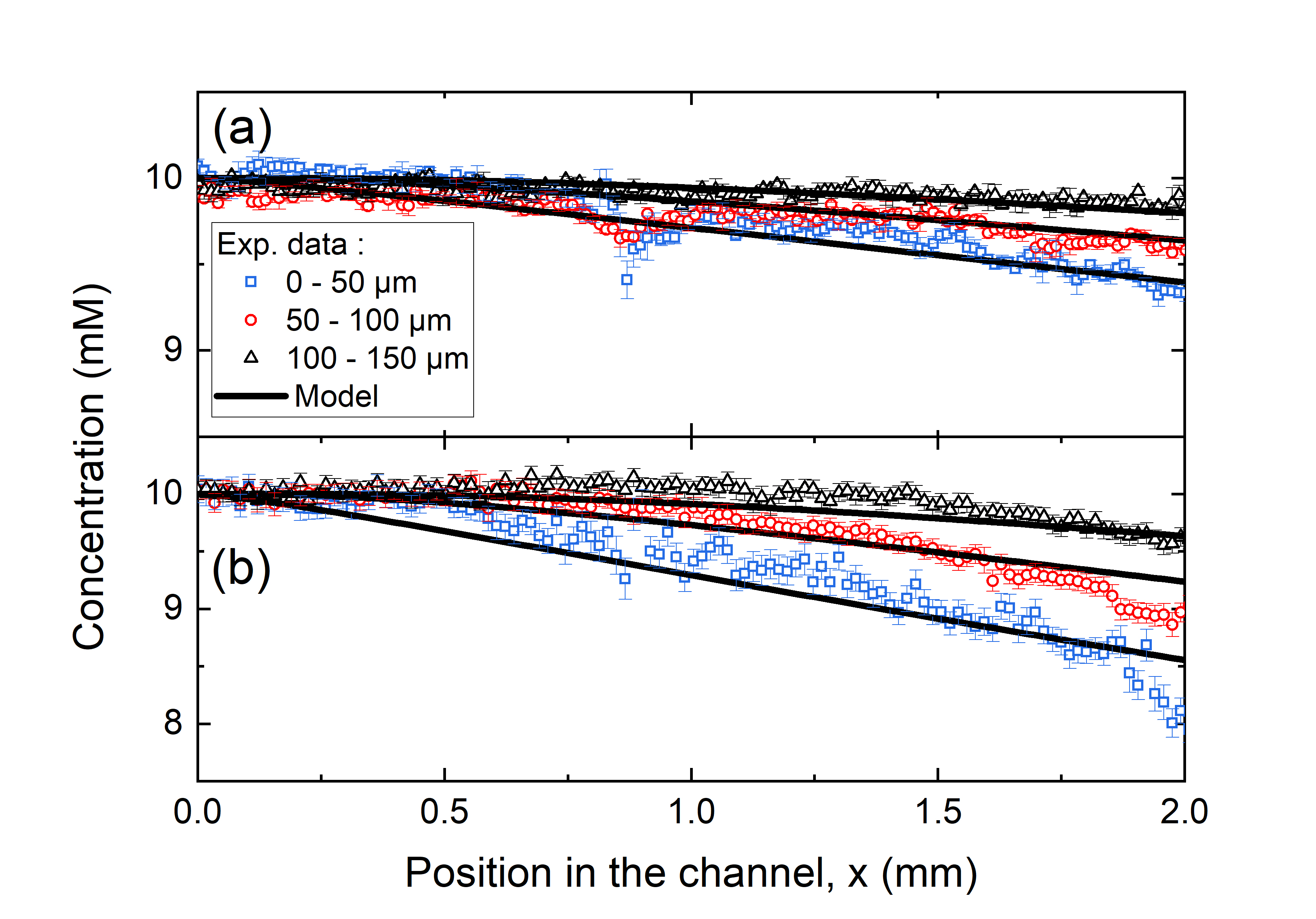}
	\caption{Comparison of the concentration distributions in the depletion zone for 3 y-positions to the analytical model once $D$ and $k_0$ were estimated. Results obtained at (a) 20 µA and (b) 40 µA. The errors bar were computed from the standard deviation of the measurements over 50 images.}
	\label{f_fig6}
\end{figure}

In this next step, the data of both depletion zones are used and averaged together. The estimation of the mass diffusivity and the kinetic reaction rate coefficient was performed with the concentration measurements in the depletion zone and the analytical model presented in section 2.3. The experimentally measured depletion zones and the analytical model's concentration distribution are presented in Figure \ref{f_fig6}.This model is only valid because of the choice in cell geometry: MFC with high aspect ratio ensuring a 2D diffusion of the concentration with an average velocity.\\

In order to increase the SNR, the concentration field in the depletion zone was averaged every {50 µm} on each side of the electrode in the y-direction (see the axes in Figure \ref{f_fig5}). In line with the experimental data, Equation \ref{e_conv_prod} was integrated between the same boundary, i.e. for the first zone : $\bar{c_1}(x)=\int_0^{l_1}c(x,y)/l_1dy$. A minimization algorithm (simplex algorithm from fminsearch function in Matlab) was then used to estimate parameters $k_0$ and $D$, minimizing the error between the model and the experimental data. The use of an analytical model enables fast processing of the data (almost real time $\sim$3 s).\\

{The identification of $D$ and $k_0$ was performed for two currents 20 and 40 µA. In Figure \ref{f_fig6}(a), a good agreement between the model and the measurements was obtained from the parametric estimation. The parameters were determined were $D=(5.5 \pm 2.5)\times 10^{-3}$ mm$^2$/s and $k_0=(0.9\pm 0.1)\times 10^{-3}$ mm/s. For a first time, an operando value of $k_0$ is provided. However, the value of the diffusivity $D$ is largely overestimated when compared to the literature. This is mainly due to the poor sensitivity of this parameter in the model used (see supplementary materials). As the sensitivity of the two parameters is quite similar, a large $D$ could be compensated by a small $k_0$ and conversely. However, the estimation procedure was still able to converge to a unique global minimum (see supplementary materials Figure S6)} and to give a correct value of $k_0$ (see next section).\\

{A second identification of the parameters was performed at  40 µA (see Figure \ref{f_fig6}(b)). It leads to $D=(3.7\pm 0.5)\times 10^{-3}$ mm$^2$/s and $k_0=(2.4\pm 0.1)\times 10^{-3}$ mm/s. The value of $D$ is still not in the range expected by the literature, but, the value of $k_0$ was increased as expected since the current density increased. Thus, the proposed method can be used to identify these parameters in simple MFC geometry. However, the correlation between both parameters would first require a precise estimation of the diffusivity, before the proposed method is used to estimate the kinetic coefficient rate. Finally, Tafel parameters could also be estimated over a range of cathode potentials, but this is out of the scope of the work, which demonstrates the use of imaging techniques to measure mass transfer parameters}.\\

\subsection{Comparison with the model in steady state}
\begin{figure}[H]
\centering
\includegraphics[scale=.5]{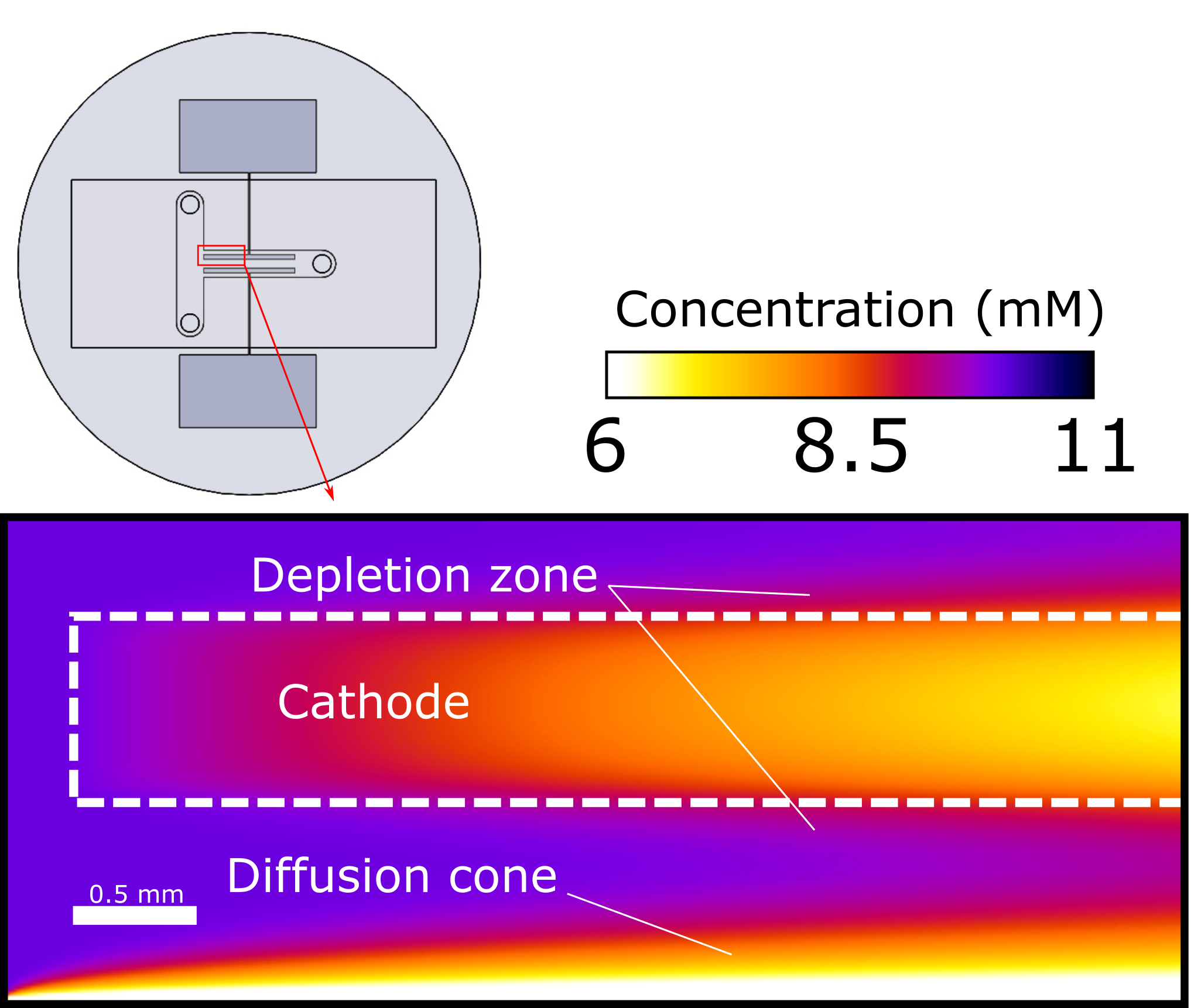}
\caption{Result of the concentration field computed using the numerical model presented in appendix. Only the electrode onset at the MFC inlet is presented to match the imaging data. The red rectangle in the MFC insert indicate the position. The white rectangle represents the electrode position.}
\label{f_fig7}
\end{figure}

In the last section, the concentration field computed using a numerical model (see appendix) is presented to validate the values of the estimated parameters. {In the numerical model, all the operating conditions and cell geometry are representative of the experiments (e.g. flow rates, current density and MFC dimensions)}. The only unknown parameters that remain are the mass diffusivity of permanganate in the aqueous solution and the kinetic reaction rate coefficient. These parameters were taken from the previous estimation at 20 µA. The resulting concentration field is presented in Figure \ref{f_fig7}.\\

Qualitatively, the concentration fields that were obtained with the numerical model are similar to the measurements presented in Figure \ref{f_fig5}, including {depletion and diffusion zones}. The magnitude is also similar, i.e. $\sim$ 2 mM in decrese for the concentration over the first 5 mm of the channel length. Another interesting result is the total current predicted by the model, which can be computed using Faraday's law (see appendix). A total current of 18.5 µA was obtained using of the estimated value of $k_0$ which is very close to the 20 µA set in the experiment. If a value of $k_0=1,04\times 10^{-3}$ mm/s was used instead (which is in the uncertainty range of this parameter), then a total current of 20 µA was obtained. This result demonstrates that with a few set of free parameters, predictive modelling for the MFC performances is viable. Further MFC characterizations {would help improve the precision of the model}.



\section{Conclusion}
In this work, an operating MFC was imaged using an in-house visible spectroscopy setup. A specific imaging setup and MFC design were specifically designed to achieve this goal. The concentration fields were measured based on the Beer-Lambert law, enabling a measurement of reactant concentration variations as low as 0.2 mM when a current is triggered. The obtained image was the first reported in the literature and shows the potential of advanced imaging techniques for MFC characterization. \\

Along with the experimental data, an analytical model of the concentration diffusion in the depletion zone close to the electrode is proposed. The comparison between this model and the experimentally meaasured concentration fields enables operando measurements of the main mass transfer parameters, i.e. $D$ and $k_0$. {Eventhough a large uncertainty in the value of $D$  was found, this values allow us to model the concentration distribution and current. Validity of the model could be enhanced by determining the diffusivity $D$ prior to the reaction rate $k_0$}.\\


Finally, this work demonstrates the feasibility of both imaging and modelling for transient concentration fields in an MFC, and the viability of this platform for characterizing the mass transfer in MFCs. {Given the importance of the numerical models in the research community for predicting the MFC performance, such characterization platforms are of prime importance for validation of this models}. However, several improvements to the present work are necessary for estimating the Tafel parameters or the impact of the operating conditions on the mass transfer parameters. {By elaborating upon the presented techniques, MFCs can be optimized as more powerful and efficient electrochemical energy conversion systems}.

\section*{Acknowledgement}
The authors gratefully acknowledge the French National Research Agency (ANR) for its support through the project I2MPAC, Grant No. ANR-20-CE05-0018-01. 

\section*{Appendix}
\subsection*{General equations of the concentration diffusion and reaction}
To simplify the model the velocity profile is assumed to be laminar, established, and modelled using a Poiseuille flow. The mass diffusivity is considered constant and the problem is solved in the steady state. Fick's law is used to model the mass diffusion in diluted solution, the ions and the reactants in the electrolyte and in the anolyte do not interact on the mass and charge transport at the cathode, and the electrochemical reaction is modelled by a Tafel law.

Under these conditions, the 3D problem of mass transport can be written as

\begin{eqnarray}
\nabla\cdot (v_xc) &=& D\nabla^2 c \label{e_transport_masse_3D} \\
\left.\frac{\partial c}{\partial y}\right|_{y=0,l_c}&=&\left.\frac{\partial c}{\partial z}\right|_{z=h}=0,\\ 
-D\left.\frac{\partial c}{\partial z}\right|_{z=0}&=&-\frac{j(x,y)}{n_eF}, \label{e_cond_flux_electrodes}\\
c(x=0,y,z) &=& c_0(y),
\end{eqnarray}
where $c$ is the reactant concentration (M); $D$ is the mass diffusivity (m$^2$/s); $x$, $y$ and $z$ are the spatial coordinates (m) as defined in Figure \ref{f_schema_3D}, with $z$ the vertical direction; $n_e$ is the number of electrons exchanged; $F$ is the Faraday constant (C/mol); and $j(x,y)$ is the current density distribution (A/m$^2$) on the electrode.The current density is zero outside of the electrode. Equation \ref{e_cond_flux_electrodes} is Faraday's law applied to the electrode interface. The current density in this equation is modelled using the Tafel law, which links the local reactant concentration to the fuel cell potential as:
\begin{equation}
j(x,y) =i_0\frac{c(x,y,z=0)}{c_0}\exp(\eta/b),
\end{equation}
where $i_0$ is the electrode exchange current (A/m$^2$), $b$ is the Tafel slope (V) and $\eta$ is the overpotential (V). The velocity profile $v_x(y,z)$ can be written analytically under the assumption of a Poiseuille velocity profile in a rectangular channel as \cite{Chevalier2021,Bruus2008}
\begin{equation}
v_x(y,z) = \frac{4h^2\Delta p}{\pi^3\mu L}\sum_{n,odd}^\infty\frac{1}{n^3}\left[1-\frac{\cosh(n\pi\frac{2y-l_c}{2h})}{\cosh(n\pi\frac{l_c}{2h})}\right]\sin\left(n\pi\frac{z}{h}\right),
\label{e_v_profile}
\end{equation}
where $h$, $L$ and $l_c$ are the channel dimensions indicated in Figure \ref{f_schema_3D}, $\Delta p$ is the pressure difference (Pa) and $\mu$ is the viscosity (Pa.s).\\
\subsection*{Equations of the numerical model used in Figure 7}
Given the geometry of the MFC, the aspect ratio of the channel, $\gamma=l_c/h$, is considered large enough to neglect diffusion in the z-direction, leading to $\partial^2c/\partial z^2\approx j(x,y)/(n_eFDh)$. The operating conditions of the MFC allows us to consider the Peclet number in the x-direction to be large enough to neglect diffusion in this direction, i.e. $Pe\gg 1$ and $\partial^2c/\partial x^2\approx 0$. The velocity components in the y-direction can also be neglected.\\
Therefore, the previous equations can be rewritten as

\begin{eqnarray}
\bar{v}_x(y)\frac{\partial \tilde{c}}{\partial x}&=& D\frac{\partial^2 \tilde{c}}{\partial y^2}-K(x,y)\tilde{c}, \label{e_transport_masse__electrode} \\
\left.\frac{\partial \tilde{c}}{\partial y}\right|_{y=0,1}&=&0,\\ 
\tilde{c}(x=0,y) &=& \Theta(y-(l_c-l_{in})/l_c),
\end{eqnarray}
where the dimensionless concentration is used, such as $\tilde{c}=c/c_0$, and $\Theta$ is the Heaviside function modelling the initial reactant concentration distribution. $\bar{v}=q_{tot}/(hl_c)$ is the average velocity of the reactants. The function $\bar{v}_x(y)$ is obtained from integrating of the velocity profile (\ref{e_v_profile}) in the z-direction as
\begin{equation}
\bar{v}_x(y) = \frac{\bar{v}}{1-0,63/\gamma}\left(1-\sum_{n,odd}^\infty\frac{96}{(n\pi)^4}\frac{\cosh(n\pi\gamma(2y-1)}{\cosh(n\pi\frac{\gamma}{2})}\right).
\end{equation}
 The function $K(x,y)$ in Equation \ref{e_transport_masse__electrode} is the kinetic rate \cite{Chevalier2021} defined as
\begin{equation}
K = \left\{
\begin{array}{lll}
 k_0/h & \mbox{if} & x,y \in \Omega_e; \\
 0 & \mbox{else} , &
\end{array}
\right.
\end{equation}
where $\Omega_e$ is the electrode domain, and $k_0= i_c^0/(n_eFc_0)e^{\eta/b}$ is the kinetics reaction rate constant (m/s).\\

The previous set of equations is solved using a numerical scheme based on Finite Difference to approximate the Laplacian in the y-direction and a Runge-Kutta algorithm in the x-direction. This numerical model is solved using MATLAB, and the subroutine \textit{ode15s} was used for the Runge-Kutta integration scheme. A total of 150 elements in the y-direction were used in the finite difference mesh, and this was considered enough to ensure mesh independence.\\

Once the concentration field is solved, the current density produced by the cell can be estimated as
\begin{equation}
I_{tot} = n_eF q_c c_0 \varepsilon,
\end{equation}
where $q_c$ is the permanganate solution flow rate, and $\varepsilon$ is the cell efficiency defined as
\begin{equation}
\varepsilon = 1-\frac{\displaystyle\int_0^{1}\tilde{c}(y,x=L_e)dy}{\displaystyle\int_0^{1}\tilde{c}(y,x=0)dy}.
\end{equation}
It is the ratio of the quantity of reactant consumed at the outlet to the initial quantity of reactant injected in the MFC at the inlet; therefore, $\varepsilon \in [0 ; 1]$. A high MFC efficiency ($\varepsilon$~1) is needed to improve this technology.

\subsection*{Analytical solution of the  concentration at the channel/electrode interface}
As mentioned in section 2.3, the concentration of the channel/electrode interface, $c_e(x) = c(y=e/2,x)$ is needed in Equation \ref{e_conv_prod}. This function is obtained using a Laplace transform of Equation \ref{e_transport_masse__electrode} for the electrode domain and for the channel domain. Moreover, a constant average velocity $\bar{v}$ also needs to be considered, which is justified in the middle of the channel regarding the aspect ratio \cite{Bazant2004}. This leads to the following equations:
\begin{eqnarray}
\frac{d^2 \hat{c}_1}{d y^2} - \alpha_1^2\hat{c}&=&\frac{k_0}{hDp}; \quad y\in[0;e/2]\\
\frac{d^2 \hat{c}_2}{d y^2} - \alpha_2^2\hat{c}&=&0; \quad y\in[e/2;\infty]
\end{eqnarray}
with 
\begin{equation}
\hat{c}_i(p)=\int_0^\infty(\tilde{c}-1)\exp(-px)dx,
\end{equation}
and $\alpha_1=\sqrt{\bar{v}p/D+k_0/(hD)}$, $\alpha_2=\sqrt{\bar{v}p/D}$, $p$ are the Laplace complex parameters, and $h$ is the channel height. These equations can be solved analytically using the adiabatic condition at $y=0$, continuity conditions at $y=e/2$ and the semi-infinity condition when $y\longrightarrow\infty$. The following expression of $\hat{c}_e(p)$ in the Laplace domain is :
\begin{equation}
\hat{c}_e(p)=-\frac{k_0}{phD\alpha_2}\frac{e^{\alpha_1e/2}\tanh(\alpha_2e/2)}{\alpha_1+\alpha_2\tanh(\alpha_2e/2)}. \label{e_ce_p}
\end{equation}
Equation \ref{e_ce_p} is then used in the inverse Laplace transform algorithm \cite{10.1145/361953.361969} to get the concentration at the channel/electrode interface for any x-position, i.e. $c_e(x)=\mathcal{L}^{-1}\lbrace\hat{c}_e(p)\rbrace$.


\end{document}